\documentclass[twocolumn,showpacs,preprintnumbers,amsmath,amssymb,prl,aps,reprin,superscriptaddress]{revtex4}
\usepackage{graphicx}
\usepackage{amsmath}
\usepackage{verbatim}

\newcommand{\ket}[1]{\left\lvert #1 \right\rangle}
\newcommand{\bra}[1]{\left\langle #1 \right\rvert}
\newcommand{\avg}[1]{\left\langle #1 \right\rangle}
\newcommand{\V}{\mathrm{V}}
\newcommand{\K}{\mathrm{K}}
\newcommand{\MHz}{\mathrm{MHz}}
\newcommand{\GHz}{\mathrm{GHz}}
\newcommand{\us}{\mu\mathrm{s}}
\newcommand{\ns}{\mathrm{ns}}
\newcommand{\fm}{f_\mathrm{m}}

\newcommand{\Real}{\mathrm{Re}}
\newcommand{\Imag}{\mathrm{Im}}
\newcommand{\VQ}{V_Q}
\newcommand{\VI}{V_I}
\newcommand{\VQbar}{\avg{V_Q}}
\newcommand{\VIbar}{\avg{V_I}}

\newcommand{\Wn}{w}

\newcommand{\Winf}{w_\infty}

\newcommand{\Wopt}{w_\mathrm{opt}}

\newcommand{\Wmm}{w_\mathrm{mm}}

\newcommand{\kJPA}{\kappa_{\mathrm{JPA}}}
\newcommand{\Rtomo}{R_{\vec{n},\overline{\varphi}}}
\newcommand{\Rsimple}{R_{\vec{n}}}
\newcommand{\Rtomofb}{R_{\vec{n},{\varphi}}}
\newcommand{\cfb}{c_\mathrm{fb}}

\newcommand{\cfbmax}{c_\mathrm{opt}}
\newcommand{\rol}{r_\mathrm{ol}}
\newcommand{\rcl}{r_\mathrm{cl}}
\newcommand{\roff}{r_\mathrm{off}}
\newcommand{\rcon}{r_\mathrm{con}}

\newcommand{\kin}{\kappa_\mathrm{in}}
\newcommand{\kout}{\kappa_\mathrm{out}}
\newcommand{\nphbar}{\bar{n}_{\mathrm{ph}}}
\newcommand{\abs}[1]{\left| #1 \right|}
\newcommand{\fQ}{\omega_{Q}}
\newcommand{\fr}{f_\mathrm{r}}

\newcommand{\Gd}{\Gamma_{\rm{d}}}
\newcommand{\avgag}[1]{\alpha_0\left(#1\right)}

\newcommand{\alphg}{\alpha_0}

\newcommand{\alphi}{\alpha_i}

\newcommand{\rhoegVint}{\rho_{01}\left(\Vint\right)}

\newcommand{\etap}{\tilde{\eta}}
\newcommand{\avgaeast}[1]{\alpha_1^*\left(#1\right)}
\newcommand{\GSD}{G_{\mathrm{s},\Delta}}

\newcommand{\bd}{b^\dagger}
\newcommand{\bdout}{\bd_\mathrm{out}}

\newcommand{\epsm}{\epsilon_\mathrm{m}}
\newcommand{\epsp}{\tilde{\epsilon}_\mathrm{m}}
 \newcommand{\MQ}{M_Q}
\newcommand{\MI}{M_I}
\newcommand{\Vint}{V_\mathrm{int}}
\newcommand{\dt}{\mathrm{d}t}

\begin{document}
\title{Reversing quantum trajectories with analog feedback}

\author{G.~de Lange\footnote[1]{These authors contributed equally to this work.}}

\author{D.~Rist\`e\footnotemark[1]}

\author{M.~J.~Tiggelman}

\affiliation{Kavli Institute of Nanoscience, Delft University of Technology, P.O. Box 5046,
2600 GA Delft, The Netherlands}

\author{C. Eichler}
\affiliation{Department of Physics, ETH Z\"{u}rich, CH-8093, Z\"{u}rich, Switzerland}

\author{L.~Tornberg}

\author{G. Johansson}
\affiliation{Department of Microtechnology and Nanoscience, MC2, Chalmers University of Technology, 
SE-412 96 Gothenburg, Sweden}

\author{A. Wallraff}
\affiliation{Department of Physics, ETH Z\"{u}rich, CH-8093, Z\"{u}rich, Switzerland}

\author{R.~N.~Schouten}
\affiliation{Kavli Institute of Nanoscience, Delft University of Technology, P.O. Box 5046,
2600 GA Delft, The Netherlands}

\author{L.~DiCarlo}
\affiliation{Kavli Institute of Nanoscience, Delft University of Technology, P.O. Box 5046,
2600 GA Delft, The Netherlands}

\date{\today}

\begin{abstract}
We demonstrate the active suppression of transmon qubit dephasing induced by dispersive measurement, using parametric amplification and analog feedback. By real-time processing of the homodyne record, the feedback controller reverts the stochastic quantum phase kick imparted by the measurement on the qubit. The feedback operation matches a model of quantum trajectories with measurement efficiency $\etap\approx 0.5$, consistent with the result obtained by postselection. We overcome the bandwidth limitations of the amplification chain by numerically optimizing the signal processing in the feedback loop and provide a theoretical model explaining the optimization result.
\end{abstract}

\pacs{03.67.Lx, 42.50.Dv, 42.50.Pq, 85.25.-j}

\maketitle

In a quantum measurement, information gain is  accompanied by backaction, altering superposition states of the observed system~\cite{Wiseman09}.
Tunable strength measurements have been devised to balance the tradeoff between information gain and backaction. 
These can be realized, for example, by controlling the interaction of the observed qubit with an ancillary qubit, followed by strong measurement of the ancilla~\cite{Pryde04,Groen13,Blok13}. 
Depending on the choice of ancilla measurement basis, the observed qubit either acquires a stochastic phase kick, or is partially projected towards one of the basis states, in a direction that is determined by the measurement result. 
Similary, a cavity mode can serve as an ancilla, with the measurement basis set by the detected field quadrature~\cite{Murch13}, and a continuous spectrum of measurement results and associated kickbacks~\cite{Hatridge13, Murch13}.

For an efficient measurement~\cite{Wiseman09}, the correlation between the stochastic evolution of the system, also known as quantum trajectory, and the measurement record of the ancilla can be exploited to undo any unwanted backaction~\cite{Wiseman95, Korotkov05} or to reverse the measurement altogether.
Probabilistic reversal of measurement backaction has been pursued  with superconducting~\cite{Katz08}, photonic~\cite{Kim09}, and ionic systems~\cite{Sherman13}. Deterministic reversal, requiring feedback control, has only been demonstrated with ions~\cite{Schindler13}. 
Recent improvements in quantum coherence in circuit quantum electrodynamics (cQED)~\cite{Devoret13} have allowed first demonstrations of feedback control with superconducting qubits. 
Digital feedback, based on fully projective measurement, enabled on-demand qubit state initialization~\cite{Riste12b, CampagneIbarcq13}, deterministic teleportation~\cite{Steffen13}, and generation of deterministic entanglement by parity measurement~\cite{Riste13b}.
Analog feedback, instead, is required to counteract the continuous spectrum of measurement kickbacks in a qubit-cavity system. A first implementation of analog feedback relied on continuous monitoring of a driven qubit to stabilize Rabi oscillations~\cite{Vijay12}. 

In this Letter, we demonstrate the real-time reversal of measurement-induced qubit dephasing in cQED, using phase-sensitive parametric amplification~\cite{Castellanos-Beltran08} and analog feedback control, as proposed in Ref.~\onlinecite{FriskKockum12}. The recovery of coherence by feedback is quantitatively consistent with a measurement efficiency $\etap\approx 0.5$ for the homodyne detection chain, closely matching the result obtained by open-loop postselection. 
Furthermore, we demonstrate a numerical procedure that finds the optimal weight function for the homodyne signal integration, circumventing the inefficiency arising from the finite detection bandwidth.

\begin{figure}
\includegraphics[width=\columnwidth]{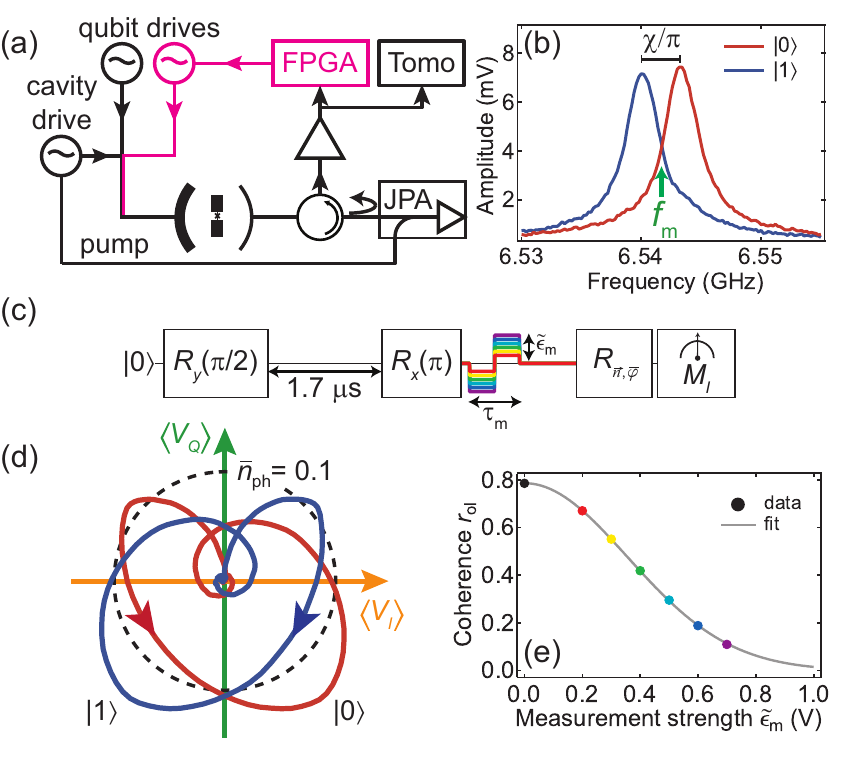}
\caption{Measurement-induced dephasing and analog feedback scheme. (a) Diagram of the key elements of the experimental setup. Qubit measurement and control drives are coupled to the input port of an asymmetrically coupled 3D cavity ($\kin/\kout \approx 1/30$). The signal emitted at the output port is added to the pump tone, which biases the JPA to a voltage gain $G=16$ and a bandwidth $\kJPA/2\pi=5.7~\MHz$ (Fig. S7). 
The reflected, amplified signal~\cite{Riste12, Castellanos-Beltran08} is 
directed by a circulator to a semiconductor amplifier (HEMT) at $3~\K$. At room temperature, the signal is split into two arms, one for data acquisition and another feeding the FPGA-based feedback controller (see Fig.~S1 for setup details). (b) Cavity spectroscopy for qubit prepared in $\ket{0}$ and $\ket{1}$. 
Measurement pulses are applied at $\fm$ (green arrow). (c) Echo sequence, where in the second half a measurement pulse with amplitude $\epsp$ is inserted to study its dephasing effect on the qubit. The second $\pi/2$ pulse is compiled into the tomographic rotation $\Rtomo$, where $\Rsimple$ is either $R_y(-\pi/2), R_x(\pi/2)$ or $I$, and the axis is rotated by $\bar{\varphi}$ around $z$ to cancel the deterministic phase shift.    
(d) Parametric plot of the averaged homodyne response $\avg{\VQ}$ versus $\avg{\VI}$ for measurement phase $\phi=\pi/2$   and $0$, respectively, for qubit in $\ket{0}$ (red) and $\ket{1}$ (blue), with $\epsp=0.4~\V$. Dashed circle: signal corresponding to $\nphbar=0.1$ intra-cavity average photon. (e) Qubit coherence $\rol$ as a function of $\epsp$. The best-fit curve gives the lever arm $\epsm/\epsp = 2\pi \times 1.2~\MHz/\V$. 
For $\epsp=0$, $\rol=\roff=0.79\pm0.01$.}
\end{figure}

We study measurement-induced dephasing of a transmon qubit (transition frequency $\fQ/2\pi=5.430~\GHz$) coupled to the fundamental mode of a 3D cavity (frequency $\fr=6.5433~\GHz$, linewidth $\kappa/2\pi=1.4~\MHz$). 
 The qubit-cavity Hamiltonian in the presence of a measurement drive at frequency $\fm$ and valid in the dispersive regime of our experiment is~\cite{Blais04}:
\begin{equation*}
\label{eq:ham}
H= (\Delta_r - \chi Z) a^\dagger a - \fQ Z/2  + \epsm(t)a +  \epsm^\ast(t) a^\dagger, 
\end{equation*}
in a frame rotating at $\fm$, with $\Delta_r/2\pi=\fr-\fm$, $a$ ($a^\dagger$) the photon annihilation (creation) operator, and $Z$ the qubit Pauli $z$-operator. 
Above, we have grouped terms to highlight the dependence of the cavity resonance on the qubit state. 
 The transmitted signal is sent to a Josephson parametric amplifier (JPA) operated in phase-sensitive mode~\cite{Castellanos-Beltran08, Riste12}. The homodyne signal obtained by demodulation is recorded for post-processing purposes and also sampled by a feedback controller implementing real-time phase correction (discussed further below) [Fig.~1(a)]. We choose for $\fm$ the average of the cavity frequencies for qubit in $\ket{0}$ ($\fr$) and $\ket{1}$ ($\fr+\chi/\pi$, with $\chi/\pi=-3.2~\MHz$)  [Fig.~1(b)].

Applying a measurement pulse entangles the qubit with the cavity field~\cite{Eichler12, Hatridge13}. 
If the measurement record is disregarded, the absolute qubit coherence $r=\abs{\rho_{01}}$ is reduced, where $\rho_{01}=\bra{0}\rho\ket{1}$ is the off-diagonal element of the qubit density matrix. 
We observe this effect by applying a pulsed measurement drive with the qubit ideally starting in the superposition state $(\ket{0}+\ket{1})/\sqrt{2}$. The measurement pulse is applied during the second half of an echo sequence [Fig.~1(c)], preferred over a Ramsey sequence to reduce the dephasing from mechanisms not inherent to the applied measurement. 
The pulse envelope has magnitude $\epsp$ and the sign reversed halfway during the total duration of $500~\ns$.  
The measured and amplified quadrature of the cavity response is set by the phase $\phi$ between the measurement pulse and the JPA pump. 
In particular, for $\phi=0$, the averaged homodyne response is equal and opposite for the qubit in $\ket{0}$ and $\ket{1}$, $\VIbar_0=-\VIbar_1$, whereas
for $\phi=\pi/2$, $\VQbar_0=\VQbar_1$  [Figs.~1(d), S2]~\cite{SOM}. 
The measurement reduces $\roff$, the qubit coherence at the end of the echo sequence for $\epsp=0$, to the open-loop coherence $\rol$. According to theory~\cite{Gambetta06}, $\rol=\roff \exp \left[-\int_0^t\Gd\left(\tau\right) d\tau\right]$, with instantaneous measurement-induced dephasing rate $\Gd(t)= 2\chi \Imag \left[\avgag{t} \avgaeast{t}\right]$, where $\alphi=\avg{a}_i$ $\propto \epsm$ is the complex-valued intra-cavity field for qubit in $\ket{i}$.  As expected, we observe a Gaussian decay of $\rol$ as a function of $\epsp$ [Fig.~1(e)]. Note that $\Gd$ is independent of $\phi$ (data not shown)~\cite{Gambetta08,Korotkov11}.

Collecting the field emitted by the cavity during a measurement reveals the quantum trajectory followed by the qubit. The measurement basis and the corresponding kickback on the qubit depend on the choice of $\phi$~\cite{Gambetta08, Korotkov11}. The $\phi$-specific  backaction becomes evident by conditioning (binning) the tomography results $\MI$ on the processed homodyne voltage. 
As first demonstrated in Ref.~\onlinecite{Murch13}, for $\phi=0$, the measurement discriminates between qubit states and coherence is lost by gradual projection to the north or south pole of the Bloch sphere (Fig.~S3~\cite{SOM}). For $\phi=\pi/2$, the case we focus on here, the measurement does not discriminate between qubit states and the kickback is a stochastic azimuthal phase $\delta \varphi$ ($z$-rotation).   
According to theory for a detector with infinite bandwidth~\cite{Gambetta08, Tornberg10, FriskKockum12}, this phase depends on the integrated weighted homodyne voltage $\Vint = \int \Wn(t) \VQ(t)  \dt$, with the weight function $\Wn(t) \propto \Real \left[ \alphg(t)\right]/\epsm$, as
\begin{equation}
\label{eq:rcl}
\rhoegVint =\roff\exp \left[(\eta-1) \int\Gd (t) \dt+i\varphi\right], 
\end{equation}
where $\varphi = c \Vint + \overline{\varphi}$, 
with $c\propto \epsm$ and $\overline{\varphi}$ the deterministic AC-Stark phase shift~\cite{Gambetta06}. Here, $\eta$ is the quantum efficiency, modeled as losses in the readout chain leading up to the JPA. In our experiment, the zero-average envelope of the measurement pulse, which makes $\int \Wn(t) \dt=0$, is chosen to suppress the infiltration of excess low-frequency noise in $\Vint$~\cite{footnotedeLange13}. Furthermore, the integration window extends  $6.5/\kappa=0.75~\us$ past the end of the applied measurement pulse [Fig.~2(a)] in order to capture the total field emitted by the cavity as it 
returns to the vacuum state~\cite{FriskKockum12}. 
Binning the tomography results $\MI$ on $\Vint$ reveals the stochastic phase $\delta\varphi$ induced by the measurement [Fig.~2(b-d)]~\cite{Murch13}. 
Rather than relying on the weight function predicted by theory, 
we numerically optimize $\Wn=\Wopt$ to maximize the conditioned coherence  $\rcon=\sum C(\Vint)r(\Vint)$, with $r$ the absolute coherence and $C$ the fraction of counts for the bin centered at $\Vint$~\cite{ SOM}. 
\begin{figure}
\includegraphics[width=\columnwidth]{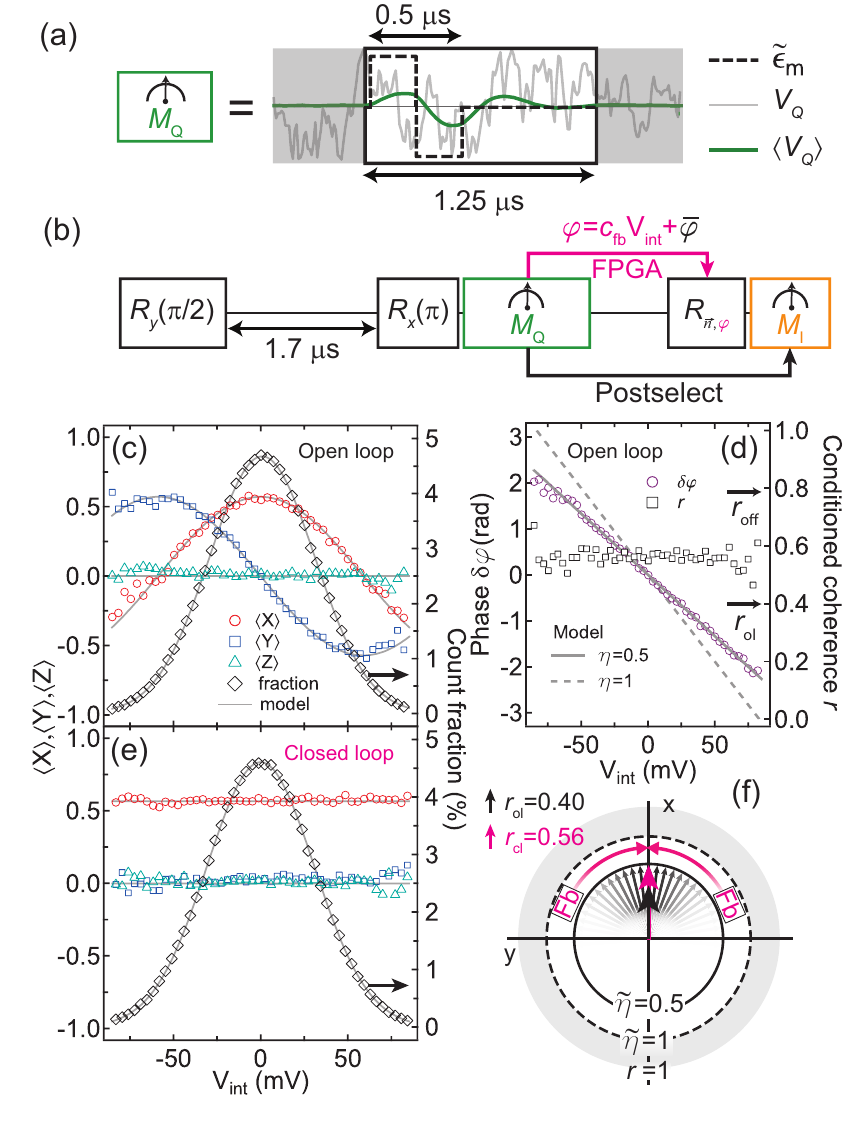}
\caption{Conditional qubit tomography and cancellation of measurement-induced dephasing by analog feedback. (a) The measurement $\MQ$ is performed with a pulse at $\fm$ with amplitude $\epsp=0.4~\V$ and $500~\ns$ length (dashed trace). The homodyne record $\VQ$ is acquired for a total duration of  $1.25~\us$ from the start of the measurement pulse. Light (dark) trace: single (average) record. (b) Measurement scheme. (c) Conditional state tomography (left) and corresponding fraction of counts $C$ (right) in open-loop operation. Solid (dashed) curves: data (model with $\etap=0.50$). The tomography outcomes $\MI$ are binned on $\Vint = \sum_{n} \Wn[n] \VQ[n]$, where $\VQ$ is sampled every $10~\ns$. The weight function $\Wn=\Wopt$ is obtained by numerical optimization using the records $\VQ$ (see also Fig.~4).  (d) Stochastic qubit phase $\delta\varphi$ (dots) and absolute coherence $r$ (squares), binned on $\Vint$, and model for $\delta\varphi$ with $\etap=0.50$ (solid) and $1$ (dashed line). In closed-loop operation [(e), corresponding to $\cfb=-10$ in Fig.~3(a)], $\VQ$  is fed to the feedback controller, which calculates $\Vint$ using $\Wopt$ and translates it into $\delta\varphi$, setting the phase of $\Rtomofb$. (f) Distribution of $\delta\varphi$  (grey scale) produced by $\MQ$ and refocusing by analog feedback (purple). 
This refocusing increases the unconditioned coherence from $\rol=0.40$ to $\rcl=0.56$. Solid (dashed) circle: $\rcl$ corresponding to the model with $\etap=0.5~(1)$.}  
\end{figure}  
From the conditioned coherence, we place a lower bound on $\eta$, absorbing signal losses after the JPA and classical processing of $\VQ$ in an overall measurement efficiency $\etap$ in Eq.~\eqref{eq:rcl}. We find quantitative agreement with the data for $\etap=0.50$ [Fig.~2(c-d)].

\begin{figure}
\includegraphics[width=\columnwidth]{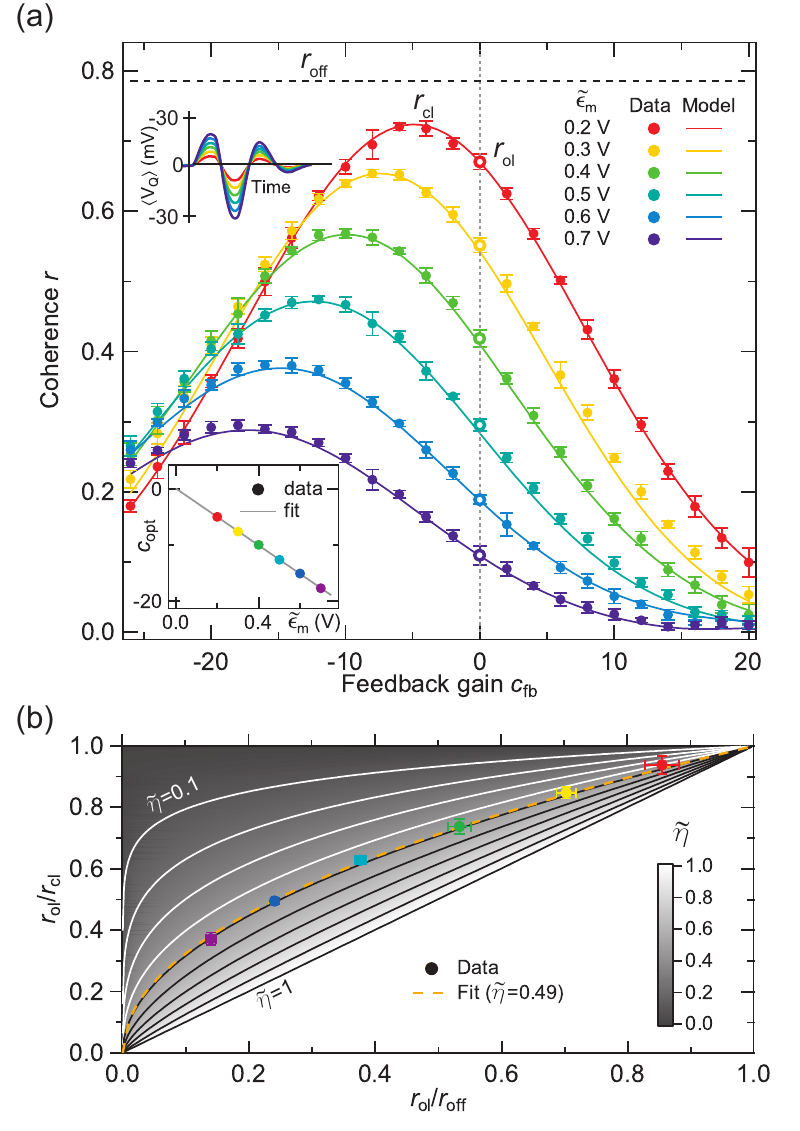}
\caption{Extraction of measurement efficiency from the extent of coherence recovery. (a) Coherence versus feedback gain $\cfb$ for $\epsp=0.2-0.7~\V$, with $\Wopt$ optimized at $\epsp=0.4~\V$. Top left: average homodyne voltage $\avg{\VQ}$ for the same range of $\epsp$. The maximum coherence $\rcl$ corresponds to the optimum feedback gain $\cfbmax$ (lower inset), directly proportional to $\epsp$. The horizontal dashed line indicates the coherence $\roff$ for no measurement drive ($\epsp = 0$). 
 Error bars are the standard deviations of $8$ repetitions. (b) Contour plot of the measurement efficiency $\etap$, with curves at $0.1$ steps. For each $\epsp$, $\rcl$ is obtained by a quadratic fit of $r$ around the maximum and $\rol$ is the measured average for $\cfb=0$ in (a). The best-fit of Eq.~\eqref{eq:eta} (orange dashed line) to the data yields $\etap=0.49 \pm 0.01$.} 
\end{figure}

Moving beyond postselection, we now set off to cancel the measurement-induced kickback  by employing analog feedback control. In real time, the controller samples $\VQ$, calculates $\Vint$ using $\Wopt$, and adjusts the phase of the tomographic pre-rotation $\Rtomofb$ by $\delta \varphi = \cfb\Vint$ (Figs.~S4, S5~\cite{SOM}). 
The optimal choice for the feedback gain ($\cfb=\cfbmax$) removes all the azimuthal phase dependence on $\Vint$ [Fig.~2(e-f)]. Crucially, $\rcon$ is unaffected, demonstrating that feedback does not introduce additional errors.   

\begin{figure}
\includegraphics[width=\columnwidth]{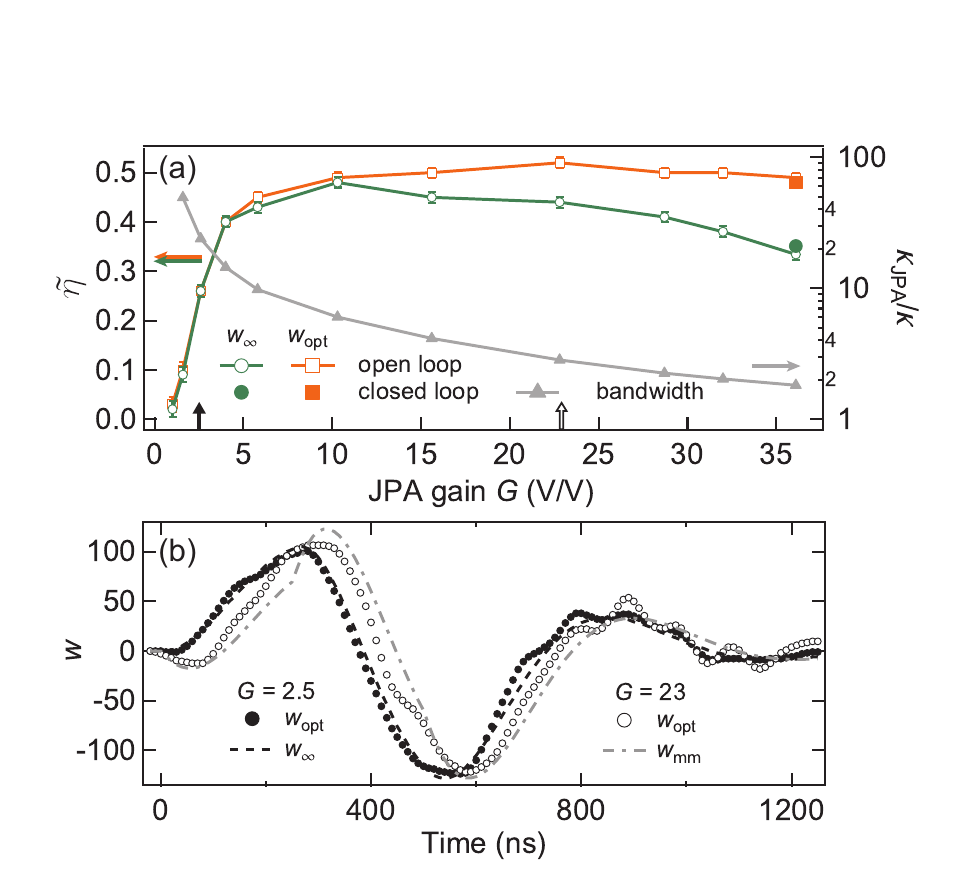}
\caption{(a) Measurement efficiency as a function of JPA gain and bandwidth. The experiment in Fig.~2(b) is repeated for different pump powers (with the JPA resonance kept at $\fm$), setting the JPA voltage gain $G$ and bandwidth $\kJPA$ (triangles). For $G\lesssim10$, the gain is insufficient to overcome the noise temperature of the HEMT (noise temperature $ \sim3~\K$). For higher $G$, the infinite-bandwidth $\Winf$ becomes suboptimal as $\VQ$ is low-pass filtered by the JPA when $\kJPA$ approaches $\kappa$ (dots). By numerically optimizing the weight function ($\Wopt$), this filtering is undone and $\etap\approx0.5$ is recovered (squares).  
For $G=1$ the pump is turned off and the JPA is intentionally detuned by $\sim200~\MHz$ from $\fm$. (b) Numerically optimized $\Wopt$ for $G=2.5$ (full dots) and $23$ (empty dots), model $\Winf$ for infinite detector bandwidth (dashed) and mode-matched $w_{\mathrm{mm}}$ (dot-dashed) for $\kJPA/2\pi=3.9~\MHz~(G=23)$.}
\end{figure}

To fully quantify the performance of the active coherence recovery, we repeat the experiment in Fig.~2(b) for various measurement-drive amplitudes $\epsp$ and feedback gains $\cfb$ [Fig.~3(a)]. 
Whereas the variance of $\Vint$ is independent of $\epsp$, as expected, the phase dependence $\mathrm{d}\delta\varphi/\mathrm{d}\Vint$ grows linearly with $\epsp$~\cite{Gambetta08,Tornberg10}, requiring the optimum $\cfbmax\propto\epsp$ [Fig.~3(a) inset]. 
Following from Eq.~\eqref{eq:rcl}, the measured $\rol$ (corresponding to $\cfb=0)$, $\roff$ $(\epsp=0)$ and $\rcl$ $(\cfb=\cfbmax)$ are related by 
\begin{equation}
\label{eq:eta}
\rol/\rcl = (\rol/\roff)^{\etap}. 
\end{equation}  
We obtain the best-fit $\etap=0.49\pm0.01$ [Fig.~3(b)].

Finally, we investigate the influence of detection settings on $\etap$. 
By adjusting the pump power, we tune the JPA voltage gain $G$ and bandwidth $\kJPA$, their product being roughly constant at $\sim90~\MHz$ (Fig.~S6)~\cite{SOM, Castellanos-Beltran08}.  
For each setting, we perform conditional tomography (as in Fig.~2) and extract $\etap$ using Eq.~\eqref{eq:eta}. 
In a first approach, we use the predicted~\cite{FriskKockum12, EichlerPhD13, SOM} weight function $\Winf\propto\Real \left[ \alphg\right]$ for infinite-bandwidth detection and unit gain
[Fig.~4(a), dots]. For decreasing gain ($G<10$), $\VQ$ is not sufficiently amplified above the noise floor of the second amplification stage at $3~\K$, causing $\etap$ to plummet. Increasing $G$ overcomes the noise floor at the expense of lowering $\kJPA$. 
However, for $G>10$, where $\kJPA \lesssim 4\kappa$, the infinite-bandwidth approximation no longer holds, resulting in a lower $\etap$.  
In a second approach, we run the numerical optimization procedure to determine $\Wopt$ at each JPA setting (Fig.~S6)~\cite{SOM}. 
In this way, we recover $\etap\approx0.5$ even as $\kJPA$ approaches $\kappa$. This independence of $\etap$ at high $G$ suggests that inefficiency arises from microwave loss between the cavity and the JPA, as assumed by the model. 
The compensation for finite detection bandwidth is reflected by the change of $\Wopt$ with $\kJPA$ [Fig.~4(b)]. 
For $\kJPA \gg \kappa$, $\Wopt$ closely matches $\Winf$. For $\kJPA \approx \kappa$, instead, $\Wopt$ differs significantly.

To understand how the JPA response impacts $\Wopt$, we apply the recent mode-matching theory of Ref.~\onlinecite{EichlerPhD13}. This theory predicts 
 the optimum weight function $\Wmm \propto \langle \bdout(t) Z \rangle$, with $\bdout(t)$ the operator for the outgoing field after amplification by the JPA~\cite{SOM}. As shown in the Supplemental Material~\cite{SOM}, $\Wmm \propto  \mathcal{F}^{-1} \left[\left(\alpha_{0,\Delta}^* - \alpha_{1,\Delta}^*\right) /2G_{\mathrm{s},\Delta}\right]$, where $\alpha_{i,\Delta} = \avg{a_{\Delta}}_i$  for qubit in $\ket{i}$, with $a_{\Delta}$ the Fourier component of the intracavity field at detuning $\Delta$ from the pump, $\GSD$ the $\Delta$-dependent small-signal gain, and $\mathcal{F}$ the Fourier transform. Interestingly, $\Wmm$ coincides with the expected $\avg{\VI}$ for qubit in $\ket{0}$, corresponding to the quadrature deamplified by the JPA   for $\phi=\pi/2$.
We find a good agreement between the predicted $\Wmm$ and the experimental $\Wopt$ [Fig.~4(b)].

In conclusion, we demonstrated the suppression of measurement-induced dephasing of a transmon qubit using parametric amplification and analog feedback. 
Optimal real-time processing of the homodyne signal makes the recovery of coherence independent of detection bandwidth and equal to the maximum achievable with the quantum efficiency $\approx 0.5$. We estimate that applying the same feedback scheme to the cavity-assisted parity measurement in the same conditions as Ref.~\onlinecite{Riste13b} would improve concurrence from the measured $34\%$ to $42\%$.  

Improving quantum efficiency will be essential to fully undo measurement kickback and for protocols, such as qubit-state stabilization~\cite{Wang01,Gillett10} and continuous-time error correction~\cite{Ahn02}, requiring near-perfect correlation between measurement record and kickback. Alternatively, analog feedback schemes that rely on qubit projection can tolerate a lower efficiency, since estimation of the quantum state improves with the measurement strength. 
Similarly to the first implementations of digital feedback in the solid state~\cite{Riste12b,CampagneIbarcq13,Steffen13,Riste13b}, which reached high fidelity in spite of moderate efficiencies, analog feedback using projective measurement offers the capability to create and stabilize entanglement~\cite{Sarovar05,Liu10} with the current state of the art.

\section{Acknowledgments}
\begin{acknowledgments}
We thank  C.~A.~Watson for experimental assistance, W.~F.~Kindel and K.~W.~Lehnert for the parametric amplifier, and A.~F.~Kockum and M.~Dukalski for helpful discussions. We acknowledge funding from the Dutch Organization for Fundamental Research on Matter (FOM), the Netherlands Organization for Scientific Research (NWO, VIDI scheme), and the EU FP7 integrated projects SOLID and SCALEQIT.
\end{acknowledgments}


\begin{thebibliography}{10}
\bibitem{footnotedeLange13}
With a square envelope, instead, we obtain $\etap=0.39$ using Eq.~(3).
\end{thebibliography}
\end{document}


\title{Supplement to ``Reversing quantum trajectories with analog feedback"}

\author{G.~de Lange\footnote[1]{These authors contributed equally to this work.}}

\author{D.~Rist\`e\footnotemark[1]}

\author{M.~J.~Tiggelman}

\affiliation{Kavli Institute of Nanoscience, Delft University of Technology, P.O. Box 5046,
2600 GA Delft, The Netherlands}

\author{C. Eichler}
\affiliation{Department of Physics, ETH Z\"{u}rich, CH-8093, Z\"{u}rich, Switzerland}

\author{L.~Tornberg}

\author{G. Johansson}
\affiliation{Department of Microtechnology and Nanoscience, MC2, Chalmers University of Technology, 
SE-412 96 Gothenburg, Sweden}

\author{A. Wallraff}
\affiliation{Department of Physics, ETH Z\"{u}rich, CH-8093, Z\"{u}rich, Switzerland}

\author{R.~N.~Schouten}
\affiliation{Kavli Institute of Nanoscience, Delft University of Technology, P.O. Box 5046,
2600 GA Delft, The Netherlands}

\author{L.~DiCarlo}
\affiliation{Kavli Institute of Nanoscience, Delft University of Technology, P.O. Box 5046,
2600 GA Delft, The Netherlands}

\date{\today}

\pacs{03.67.Lx, 42.50.Dv, 42.50.Pq, 85.25.-j}

\maketitle

\section{Device parameters}
We used the same device as in Ref.~\onlinecite{Riste13b}. The experimental parameters  differ slightly due to a different choice of cavity couplings and aging of the qubit junctions during the thermal cyclings.  Throughout the experiment, only the single-junction transmon is used and the double-junction qubit is parked at the $7.689~\GHz$ sweet spot.  
 The copper cavity has fundamental mode $\fr=6.5433~\GHz$ and coupling limited linewidth $\kappa/2\pi=1.4~\MHz$, with asymmetric coupling $\kout/\kin \approx 30$. The single-junction transmon has transition frequency $\fQ/2\pi=5.430~\GHz$, relaxation time $\Tone=27~\us$, Ramsey time $\Ttwostar=5~\us$, and echo time $\Techo=8~\us$.  The residual excitation of $\sim1\%$  is neglected in the analysis. 

\section{Analog feedback for phase cancellation}
The analog feedback loop consists of a FPGA-based controller, a voltage-controlled delayed trigger, and an arbitrary waveform generator (Tektronix AWG520). Another arbitrary waveform generator (Tektronix AWG5014) provides all the deterministic qubit control and measurement pulses and synchronizes the feedback-loop components. Measurement pulses, with carrier frequency $\fm$ and envelope $\epsm(t)$, are applied to the cavity input. The output signal is amplified and demodulated to produce the homodyne voltage $\VQ$, constituting the input to the feedback loop.  
The feedback controller digitizes $\VQ$ at $100~\MSample$ and $8$-bit resolution. The digitized signal is weighed in real time by a sequence $\Wn$ of $7-$bit signed integers, generating a running integral over $1.25~\us$. The resulting weighted integral $\Vint$ is multiplied by an integer $\cfb$, setting the analog feedback strength. 

Following digital-to-analog conversion, $\cfb\Vint$ provides the input to the delayed trigger. Upon activation by a marker bit from the AWG5014, this device starts ramping an internal voltage linearly (see Fig.~S5). A trigger is generated when the running voltage crosses $\cfb\Vint$. The trigger delay determines the phase correction for the measurement-induced phase shift and is here compiled into the tomographic pulse. 
This correction is applied by time-shifting the envelope of the tomographic pulse modulating the qubit drive tone. The use of single-sideband modulation translates this delay into a difference $\delta \varphi$ in the rotation axis in $\Rtomofb$ [Fig.~2(b)]. The modulation frequency of $30~\MHz$ achieves a phase resolution of $\sim10$~degrees, set by the $1~\ns$ clock of the AWG520. This discretization corresponds to an error of $\sim0.1\%$  in $\rcl$.

\section{Pump leakage suppression}
To ensure that the qubit does not suffer unintended measurement-induced dephasing when $\epsm=0$, the cavity needs to be empty at steady state.   
This requires cancelling the leakage of the JPA pump towards the cavity. The three circulators between cavity and JPA provide $\sim70~\dB$ suppression, but additional $20~\dB$ are desirable to prevent unwanted dephasing. 
To suppress the residual leakage, we supply a continuous-wave tone at $\fm$, by fine-tuning the DC offsets at the corresponding mixer. To calibrate these offsets, we integrate the homodyne voltage before and after applying a $\pi$ pulse to the qubit. If there are no photons in the cavity, the signal remains unchanged. In the presence of a residual photon population, instead, the transient of the intra-cavity field from the one corresponding to qubit in $\ket{0}$ to the one for $\ket{1}$ produces  a variation in the homodyne signal. We optimize the amplitude and phase of the input offset by minimizing the variance of the homodyne voltage over this interval. From the magnitude of the applied offsets, we estimate a pump leakage of $\sim10^{-2}$ intra-cavity photons without nulling. Although complete cancellation from the cavity input port cannot be achieved, due to the asymmetry in input and output coupling rates, we estimate from Fig.~S3(c) that the pump leakage is suppressed to better than $10^{-4}$ photons. This is at least two orders of magnitude lower than the steady-state population at $\epsm$ values used in the experiment.

\section{Weight function optimization}
We obtain the weight function $\Wopt$ by the following numerical optimization procedure. The results of $200,000$ pairs of experimental homodyne records $\VQ$ and tomographic measurements $\MI$ for each tomographic pre-rotation [Fig.~2(d)] are stored and processed to calculate $\rcon$ for each $\Wn$. Fig.~S7 depicts the optimization procedure starting from $\Wn=0$.  The optimization routine randomly selects one of $25$ blocks $\Wi$ (each $50~\ns$ long) of $\Wn$. Every $\Wi$ is stepped across $5$ values, while keeping the remaining blocks fixed. A quadratic fit of  $\rcon$ selects the optimum $\Wi$ at each iteration. The optimization over the whole integration window is repeated with increasingly smaller steps for $\Wi$.  
To speed up the optimization procedure for the $\Wopt$ used in Figs.~2-4 and S2, the optimization starts from $\Wn = \Real \left[\alphg\right]$, the expected optimum for a detector with infinite bandwidth~\cite{FriskKockum12} and is repeated three times. 
The final shape is obtained after final linear interpolation between adjacent $\Wi$ and smoothing by averaging each value $\Wn[n]$ with its nearest neighbors (repeated twice).

\section{Mode-matching theory for the optimal weight function}

In order to obtain the maximum correlation between measurement record and qubit state, we here apply the method described in Ref.~\onlinecite{EichlerPhD13} 
 to derive the state-dependent field propagating in the coax line from the cavity to the JPA, where it is reflected and amplified. 
Applying a measurement pulse to the cavity, with the qubit in a superposition state,  entangles qubit and cavity due to their dispersive  interaction [Eq.~(1)].
The intra-cavity field mainly exits the output port at a rate $\kappa_\mathrm{out}\approx \kappa$. The qubit-cavity interaction is complete when the cavity has returned to the vacuum state, i.e.,  several cavity decay times $1/\kappa$ after the measurement pulse is turned off. The entanglement is now between the qubit and the outgoing field $\aout$ in the coax line. 
Input-output theory~\cite{Gardiner04} connects this field to the incoming field $\ain$ and the field $a$ inside the cavity: 
\begin{equation}
\label{eq:iotheory}
\aout = \sqrt{\kappa}a -\ain.
\end{equation}
Ideally, the incoming field is in the vacuum state $\ket{\mathrm{vac}}$. The field $\aout$ is the input to the JPA ($\bin = \aout$) and is transformed to the outgoing field $\bout$. 
A time-independent field can be defined by integrating the time-dependent $\bout(t)$:
\begin{equation*}
B = \int w (t) \bout(t)  \dt,  
\end{equation*}
where $w (t)$ 
is normalized  
to preserve the commutation relation $\left[B,B^\dagger\right]=1$.  
The joint qubit-field state is now $\ket{\Psi} = \left(\ket{0}\ket{\betag, G_{\phi}e^{i \phi}} + \ket{1}\ket{\betae,G_{\phi}e^{i \phi}}\right)/\sqrt{2}$, 
where the squeezed state $\ket{\betai,G_{\phi}e^{i \phi}} = D(\betai)S(G_{\phi}e^{i \phi})\ket{\mathrm{vac}}$ is defined by the displacement operator $D(\betai) = \exp(\betai B^\dagger - \betai^\ast B)$ and the squeezing operator $S(G_{\phi}e^{i\phi}) = \exp(G_{\phi}e^{-i\phi}B^2- G_{\phi}e^{i\phi}B^{\dagger2})$~\cite{Gardiner04}.  
Here, $\betai = \avg{B}_i $ for qubit in $\ket{i}$ and $G_{\phi}e^{i\phi}$ is the complex-valued amplitude gain of the JPA, for the quadrature with phase $\phi$. 
In our experiment $\phi = \pi/2$, resulting in amplification of the $Q$-quadrature.  
The entanglement between the field $B$ and the qubit is maximized when the distance $\abs{\betag - \betae}$ is largest. This condition is matched for $w(t)=\Wmm(t) = N \langle \bdout(t)Z\rangle$~\cite{EichlerPhD13}. When the qubit starts in a maximal superposition state, this gives  
\begin{equation*}
\abs{\betag - \betae}_{\mathrm{max}}/2 = N \int  \abs{\langle \bout(t)\rangle_0-\langle \bout(t) \rangle_1 }^2\dt,
\end{equation*}   
with the normalization constant 
\begin{equation*}
N=\frac{1}{\sqrt{\int|\langle \bdout(t)\rangle_0-\langle \bdout(t) \rangle_1|^2\dt}}. 
\end{equation*}

To determine $\Wmm (t)$ for finite JPA bandwidth we move to the frequency domain, where 
\begin{equation}
\label{eq:bout}
\boutD = G_{\mathrm{s},\Delta} \binD + G_{\mathrm{i},\Delta} \bdinmD,
\end{equation}
with $G_{\mathrm{s},\Delta}$ and $G_{\mathrm{i},\Delta}$ complex gain factors. 
Throughout the text, we refer to $\abs{G_{\mathrm{s},0}}$ as the JPA voltage gain $G$.
In the small-signal approximation~\cite{EichlerPhD13} and with the pump resonant with the JPA, 
\begin{align}
\label{eq:complexgain}
G_{\mathrm{s},\Delta} &=& -1 + \frac{\kappae\left({\kappae}/{2}-i\Delta\right)}{\left(i\Delta - \lambda_- \right)\left(i\Delta - \lambda_+ \right)}\\
G_{\mathrm{i},\Delta} &=& \frac{\mathcal{G}-1}{\mathcal{G}} \frac{\kappae^2/{2}e^{-i2\phi}}{\left(i\Delta - \lambda_- \right)\left(i\Delta - \lambda_+ \right)},
\end{align}
with $\kappae$ the extrinsic loss rate of the JPA and $\lambda_{\pm} = \kappae/2 \left[1 \pm (\mathcal{G}-1)/\mathcal{G}\right]$, where $\mathcal{G}$ is set by the pump power. 
For the amplified $Q$-quadrature (for $\phi = \pi/2$) the peak gain is $G_{Q}=\abs{G_{\mathrm{s},0}-G_{\mathrm{i},0}} =  2(\mathcal{G}-1/2)$. For the deamplified $I$-quadrature, $G_{I}=\abs{G_{\mathrm{s},0}+G_{\mathrm{i},0}} =  \left[2(\mathcal{G}-1/2)\right]^{-1}$. 
Increasing $\mathcal{G}$ reduces the detection bandwidth as $\kJPA \approx \kappae/\mathcal{G}$ (Fig.~S6).
Combining Eqs.~\eqref{eq:iotheory} and ~\eqref{eq:bout}, we obtain the expectation values
\begin{align*}
&\langle{\bdoutD}\rangle_0 = \sqrt{\kappa}\left[\GcSD \avgacg{\Delta}+\GcID \avgag{-\Delta} \right] \\
&\langle{\bdoutD}\rangle_1 = \sqrt{\kappa}\left[\GcSD \avgace{\Delta}+\GcID \avgae{-\Delta} \right]. 
\end{align*}
Using $\avgag{\Delta} = -\avgace{-\Delta}$, valid for our choice of $\fm$, we arrive to
\begin{align*}
\langle{\bdoutD Z}\rangle &= (\langle{\bdoutD}\rangle_0 - \langle{\bdoutD}\rangle_1)/2\\
&=\sqrt{\kappa}\left(\GcSD +\GcID\right)\left[\avgacg{\Delta}-\avgace{\Delta}\right]/2.
\end{align*}
and $\Wmm(t) = N \mathcal{F}^{-1} \left[\langle{\bdoutD Z}\rangle\right]$, with $\mathcal{F}$ the Fourier transform.  Thus, $\Wmm$ is proportional to the average deamplified $I$-quadrature, $\Real  \left[\avg{\bout(t)}_0 \right]$. In the limit $G \gg  1$ and for $\phi = \pi/2$, $ \GcSD +\GcID \approx \left(2\GSD\right)^{-1}$. 
For $G=1$, $\Wmm(t) \propto \avgag{t} - \avgae{t} = 2\Real \left[\avgag{t}\right]$, corresponding to the $\avg{\VI}$ that would be measured in the absence of the JPA, reproducing the result in Ref.~\onlinecite{FriskKockum12}. 

\begin{figure*}
\includegraphics[width=1.3\columnwidth]{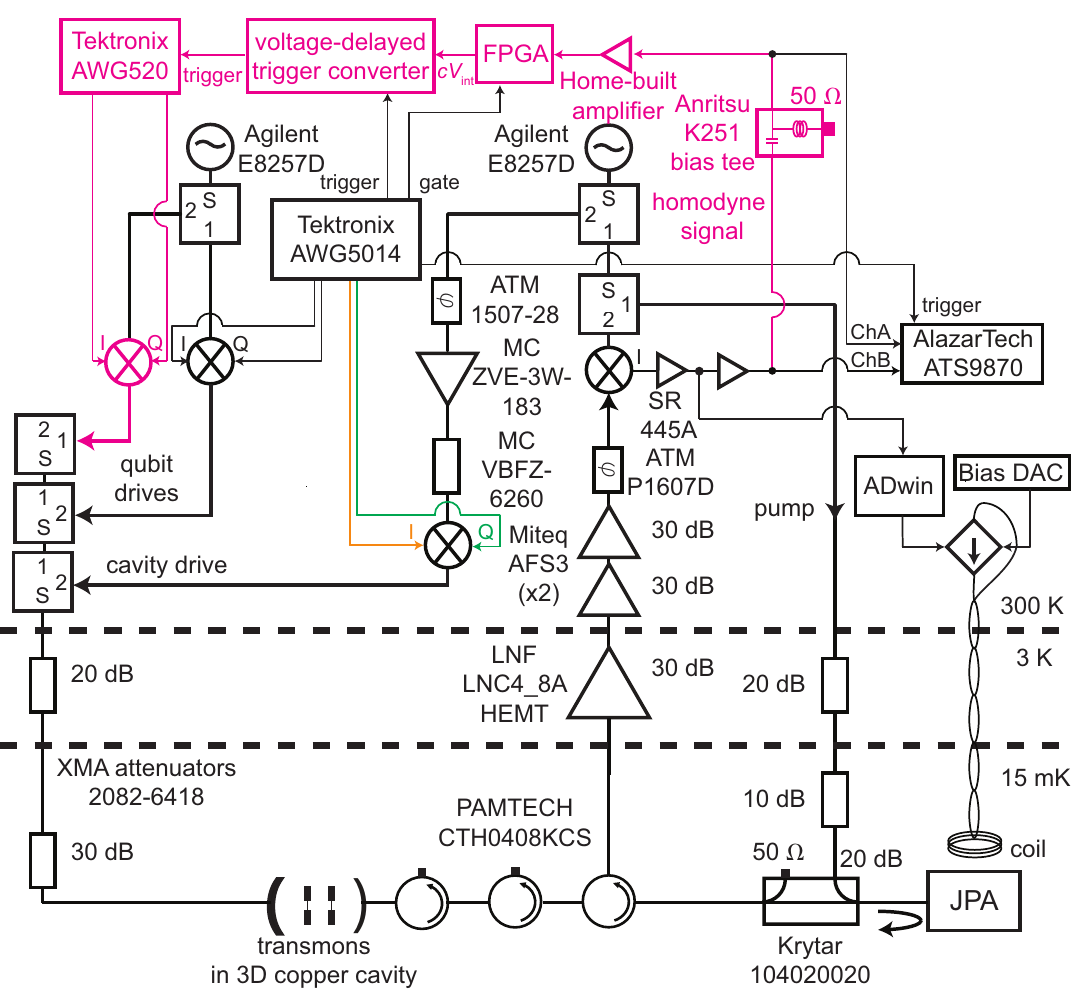}
\caption{Detailed schematic of the experimental setup. The components of the analog feedback loop are highlighted in purple. The homemade feedback controller is based on an FPGA board (Altera Cyclone IV), programmed with the weight function $\Wn$. The integrated, rescaled homodyne signal $\cfb\Vint$ is fed to a voltage-controlled delayed trigger. The signal-dependent delay translates into a phase shift in the tomographic pulse, generated by a Tektronix AWG520 modulating the qubit drive generator (see text). Qubit rotations (both deterministic and conditional) are Gaussian DRAG pulses~\cite{Motzoi09} ($\sigma=6~\ns$, $24~\ns$ total duration), with $30~\MHz$ single-sideband modulation. The homodyne signal is offset-subtracted by a bias tee and amplified in multiple stages (including a home-built amplifier with $40~\dB$ gain, $100~\MHz$ bandwidth) before entering the FPGA to span most of the fixed ADC input range ($-1$ to $1~\V$).
Other system components (black) are described in Ref.~\onlinecite{Riste12}.} 
\end{figure*}

\begin{figure}
\includegraphics[width=\columnwidth]{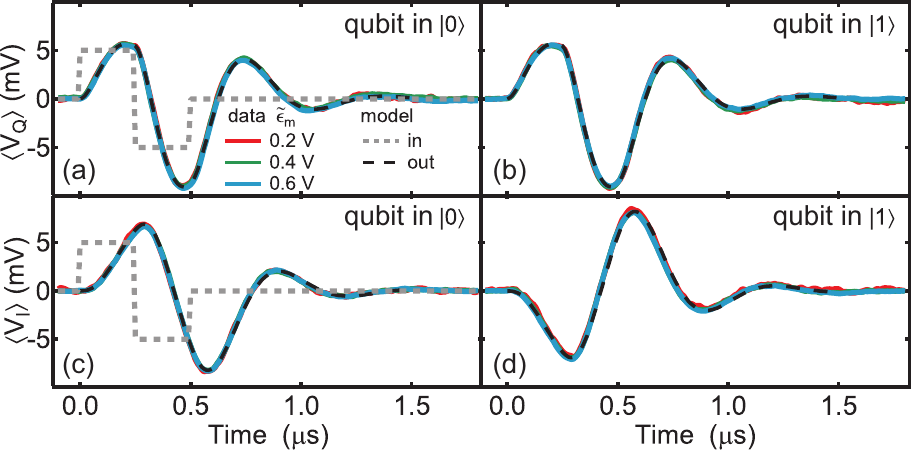}
\caption{Linearity of homodyne voltage. Averaged homodyne response ($100,000$ repetitions) for various measurement drive amplitudes $\epsm$ and homodyne detection phases $\phi=\{0,\pi/2\}$.  The qubit is prepared in either $\ket{0}$ (a,c) or $\ket{1}$ (b,d) and the applied measurement pulse is phase-shifted by either $\phi=\pi/2$ (a,b) or $\phi=0$ (c,d) relative to the pump. The excellent overlap between all curves, rescaled  by $\epsp/0.2~\V$, demonstrate the linearity of the JPA in the operating regime and evidence near-perfect agreement with the model.}
\end{figure}

\begin{figure*}
\includegraphics[width=2\columnwidth]{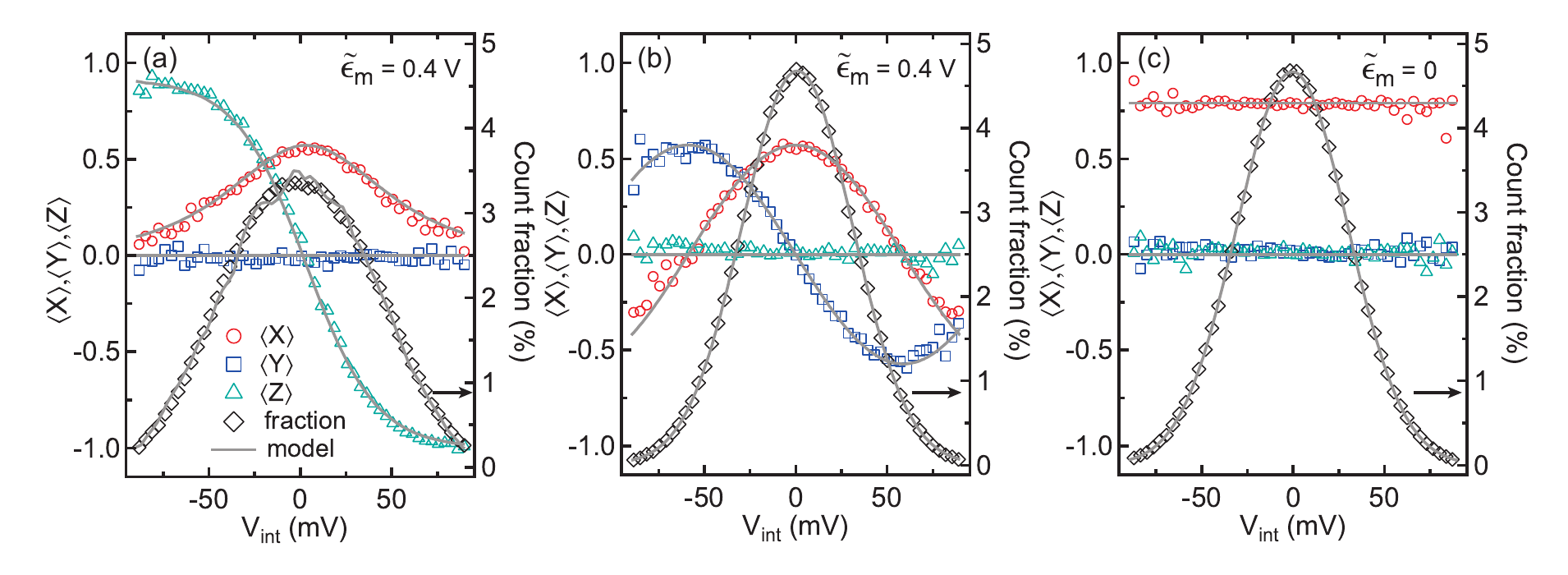}
\caption{Conditional state tomography for various measurement configurations. (a) $\phi=0$, giving maximum discrimination between qubit in $\ket{0}$ and in $\ket{1}$. (b) $\phi=\pi/2$, replicating Fig.~2(d). (c) No measurement pulse. The observed independence of the qubit state on $\Vint$ shows that any residual pump leakage into the cavity is negligible. In the three cases, $\Vint$ is calculated using the numerically optimized $\Wopt$ for (b). Panels (a-b) connect with the experimental results first shown in Ref.~\onlinecite{Murch13}.}
\end{figure*}

\begin{figure*} 
\includegraphics[width=1.7\columnwidth]{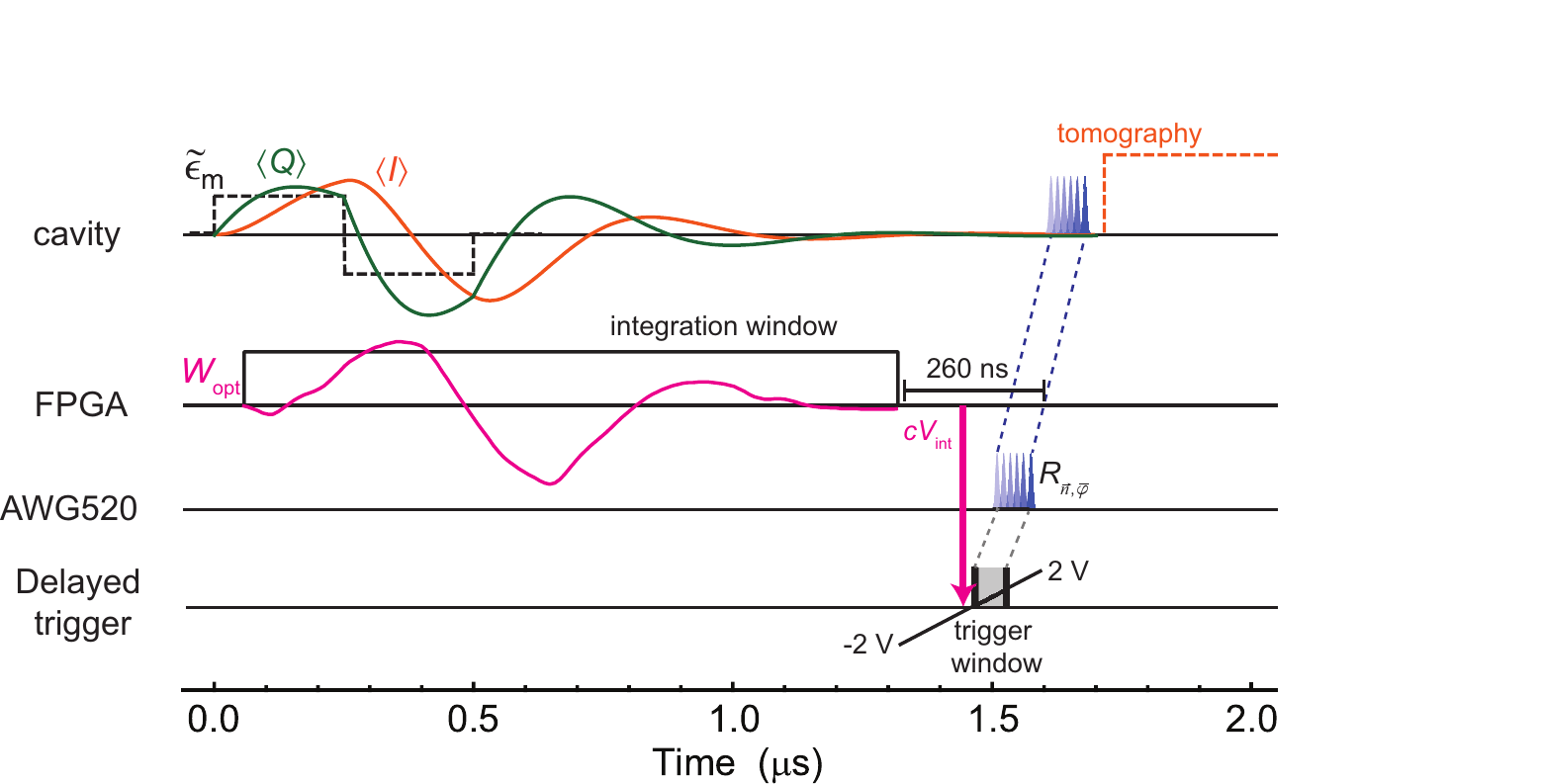}
\caption{Timings in the feedback scheme. Time $t=0$ corresponds to the end of the $\pi$ pulse in the echo sequence. From top to bottom, first row: intra-cavity quadratures $\avg{I}$ and $\avg{Q}$ upon microwave excitation at $\fm$ with pulse envelope $\epsm$, followed by conditional tomographic pulse and final readout. Second row: FPGA integration window, with weight function $\Wopt$ and output voltage $\cfb\Vint$. Third row: tomographic pulses generated by an AWG520, and delayed by a voltage-controlled trigger (fourth row, Fig.~S5). The feedback latency, defined as the time between the end of the FPGA integration and the earliest tomographic pulse reaching the cavity, is $260~\ns$.}
\end{figure*}

\begin{figure*}
\includegraphics[width=1.3\columnwidth]{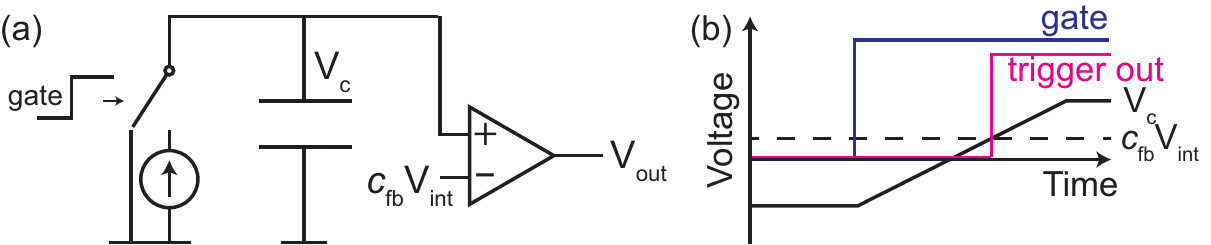}
\caption{Voltage-controlled delayed trigger. (a) Simplified diagram of the homemade circuit. 
A current source, switched by a marker bit from the AWG5014, ramps a voltage $V_C$ linearly from $-2$ to $2~\V$ in $200~\ns$. This voltage is compared to the signal from the FPGA, $\cfb\Vint$. (b) The output trigger is generated when $V_c$ crosses $\cfb\Vint$.}
\end{figure*}

\begin{figure}
\includegraphics[width=0.75\columnwidth]{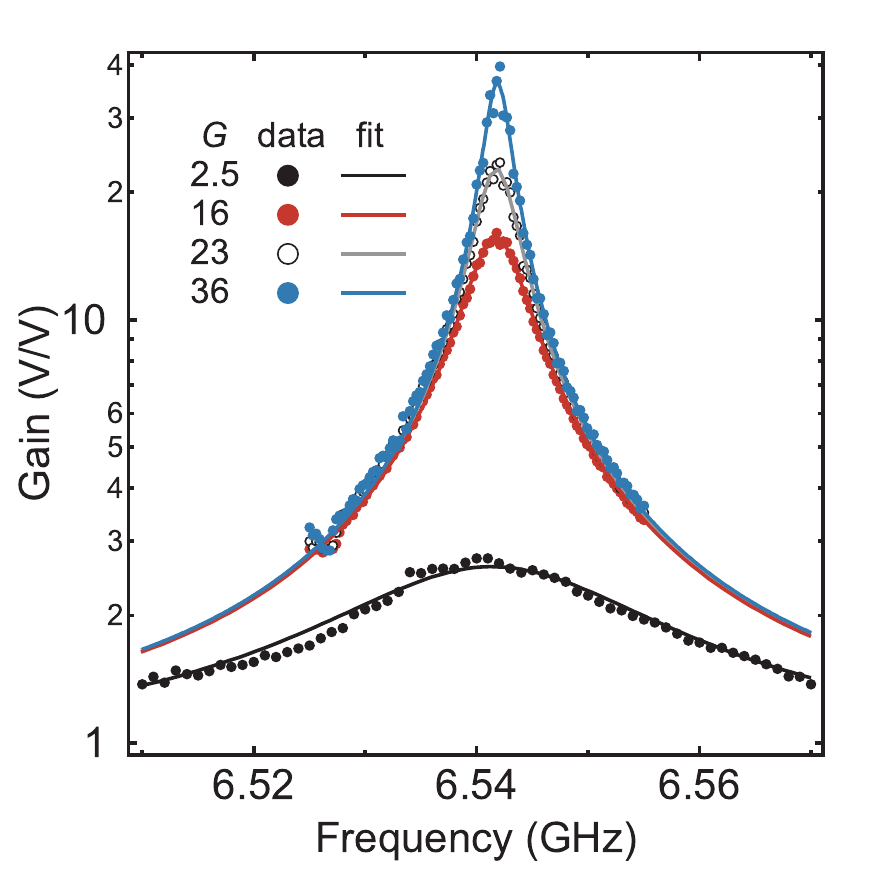}
\caption{JPA small-signal amplitude gain at different bias points. 
The values of $G$ are obtained by fitting $\abs{\GSD}$ to each curve using Eq.~\eqref{eq:complexgain}. The other fit parameter is $\kappae/2\pi = 83, 91, 92$ and $91\pm1~\MHz$ for $G=2.5, 16, 23$ and $36$, respectively. As expected, $\kappae \approx \mathcal{G}\kJPA$ is approximately constant for $G\gg1$. }
\end{figure}

\begin{figure*}
\includegraphics[width=1.8\columnwidth]{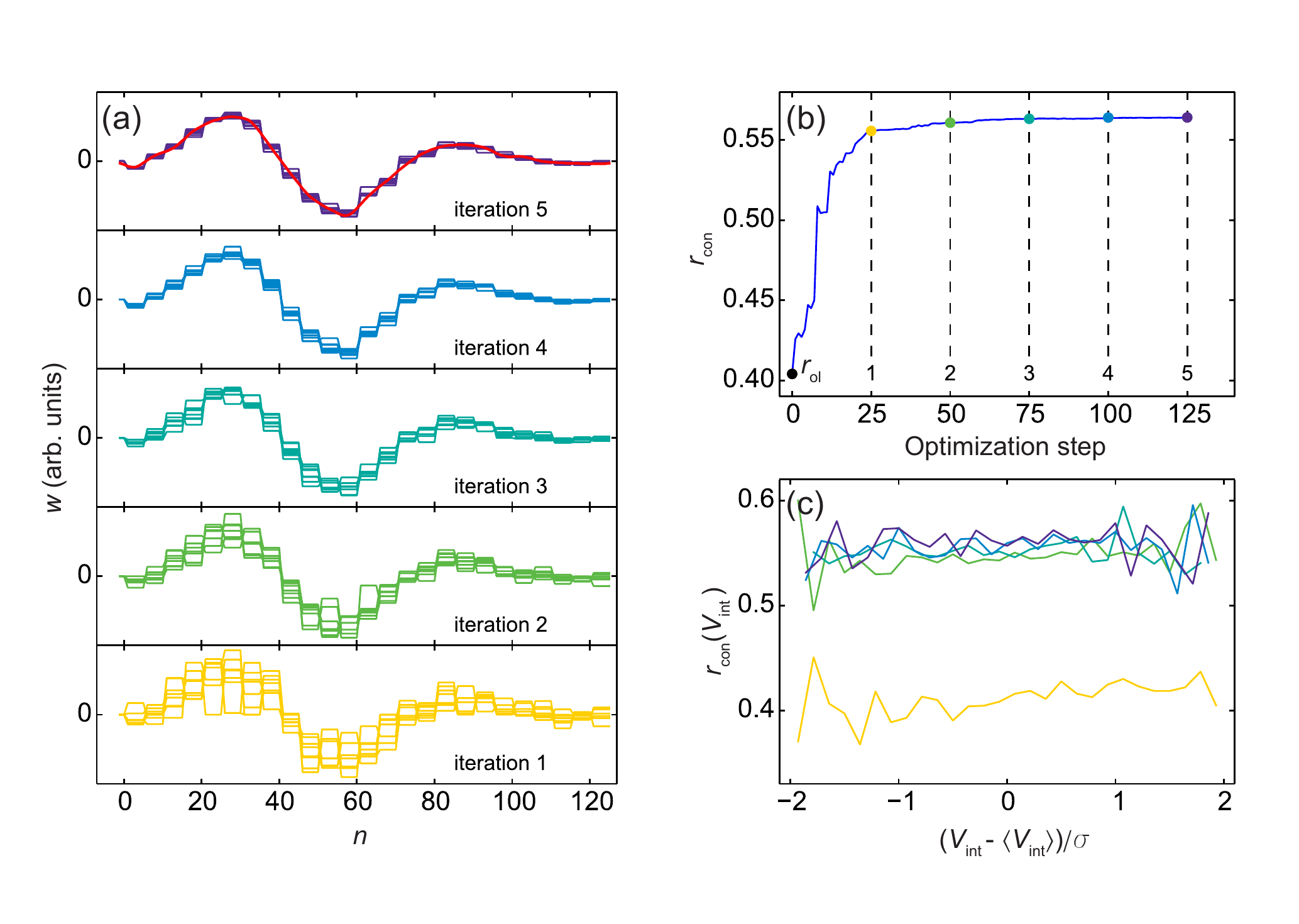}
\caption{Weight function optimization. (a) Five consecutive iterations of the optimization procedure for $\kappa/2\pi=5.7~\MHz$.  For each iteration, every $\Wi$ is stepped across seven values, with step size decreasing with the iteration number as $\left[1\colon1\colon0.8\colon0.8\colon0.5\colon0.2\right]$. 
To illustrate the convergence, the whole procedure is repeated eight times, each time starting from $\Wn=0$. The results of each iteration are superimposed.  
In every case, the result of the optimization converges to $\Wopt$ used in Figs.~2, 3, and S2 after smoothing (red curve).  
(b) $\rcon$ after each optimization step in $\Wi$, with 25 steps per iteration (dashed lines), shown for one of the optimization runs shown in (a). (c) Conditioned $r$ on $\Vint$ for successive iterations of the optimization run shown in (b). The horizontal axis is rescaled by the standard deviation $\sigma$ of $\Vint$.}
\end{figure*}